\documentclass[reqno]{amsart}

\usepackage{amsfonts}
\usepackage{amsmath}
\usepackage{amssymb}
\usepackage{mathptmx}
\usepackage[all]{xy}

\newtheorem{remark}{Remark}{\upshape\bfseries}{\upshape}

\newcommand{\comp}{\mbox{\small $\circ$}}

\let\bb\mathbb

\title{Remarks on the Configuration Space Approach to
Spin-Statistics}

\author{Andr\'es F. Reyes-Lega and Carlos Benavides}
\thanks{\hspace{-0.45cm} \vspace{0.2cm}\\Andr\'es F. Reyes-Lega\\
Departamento de F\'isica, Universidad de los Andes, Bogot\'a,
Colombia\\
E-mail: anreyes@uniandes.edu.co\\
\vspace{0.05cm}\\
\hspace{-0.45cm}Carlos Benavides\\
Departamento de F\'{i}sica, Universidad de los Andes, Bogot\'a, Colombia\\
              \emph{Present address:} Fondo Nacional de Garant\'ias. Cr. 13 No. 32-51, Bogot\'a, Colombia\\
                  E-mail: carlos.benavides@fng.gov.co
}

\begin{document}
\begin{abstract}
The angular momentum operators for a system of two spin-zero indistinguishable particles are constructed,
using Isham's Canonical Group Quantization method. This mathematically rigorous method provides a hint at the
correct definition of (total) angular momentum operators, for arbitrary spin, in a system of
indistinguishable particles. The connection with other configuration space approaches to spin-statistics is
discussed, as well as the relevance of the obtained results in view of a possible alternative proof of the
spin-statistics theorem.
\end{abstract}
 \maketitle
\section{Introduction}
\label{sec:introduction} The interest in a better understanding of the spin-statistics connection has
increased in the last years. In particular, many ideas having as final aim a derivation of the connection
within non-relativistic quantum mechanics have been
discussed~\cite{Balachandran1993,Berry1997,Anastopoulos:02,Kuckert:04,Papadopoulos2004,Harrison2004,Peshkin2003,Peshkin2006}.
In this context, the properties of the configuration space $Q_N$  of $N$ identical (indistinguishable)
particles have been particularly emphasized, in view of possible physical implications. The main fact
motivating this approach is well known: In three spatial dimensions, the Fermi-Bose alternative emerges
naturally (for spin zero particles) from the topology of the configuration space. In two spatial dimensions,
the possibility of anyonic statistics also follows from the topological properties of the corresponding
configuration space. Thus, by including the indistinguishability of quantum particles at the configuration
space level, the need for a further ``symmetrization postulate'' disappears. The history of the developments
associated to this circle of ideas is quite complex and involves the contributions of many authors. This
includes the implementation of Feynman's path integral approach on multiply-connected spaces as initiated by
Schulmann~\cite{Schulman:68}, its application to systems of identical particles by Laidlaw and
DeWitt~\cite{Laidlaw1971} and the more geometric formulation of Leinaas and Myrheim~\cite{Leinaas1977}. For a
more detailed description of these (and more recent) developments, the reader is referred to
\cite{Reyes2006,Benavides2009,Reyes-Papadopoulos2009} and to the references cited therein. In this paper, our
main purpose is to draw attention to a point that, in our opinion,  appears not to have been taken
sufficiently into account, namely, the relevance of the algebra of operators for any quantum description of a
system of identical particles based on the configuration space $Q_N$. The subtleties involved in the correct
definition of infinitesimal (i.e. self-adjoint) generators of symmetries, their domains and the algebras they
represent are well known in the context of quantization theory. But, in our view, in the context of the
non-relativistic approach to spin-statistics they have received very little attention. We believe that a
careful analysis of the representation-theoretic and functional-analytic issues involved in such an approach
could, eventually, shed new light into the problem.

Having this aim in mind, in the present paper we will, as a first step in that direction, consider the
construction of the infinitesimal generators of rotations for a system of two indistinguishable spin zero
particles. The construction is based on Isham's canonical group quantization~\cite{Isham1984}. The results
obtained will allow us to establish contact with other approaches and hence to discuss their physical
meaning. Additionally, in this paper we will show that, contrary to the opinion of some authors (see, for
example, \cite{Sudarshan2003,Allen2003,Streater:homepage}), constructions like the one developed by Berry and
Robbins~\cite{Berry1997}, although different in some crucial respects to the usual form of quantum mechanics,
do perfectly fit into a general scheme of quantum mechanics, based on the generally accepted postulates.

The paper is organized as follows. In section \ref{sec:ccr}, some well known facts about canonical
commutation relations are reviewed. With this motivation, we briefly review Isham's Canonical Group
Quantization method in section \ref{sec:isham}, emphasizing (i) the construction of unitary representations
of the canonical group for homogeneous spaces and (ii) the role played by the fundamental group of the
configuration space and its associated universal cover fibration. In section \ref{sec:RP2} we then proceed,
using Isham's method, to exhibit the canonical group for a system of two indistinguishable particles in 3
spatial dimensions. The unitary representations of the canonical group are used to explicitly construct the
infinitesimal generators of rotations for this system. Section \ref{sec:other} contains the main results of
the paper. There  we establish explicit connections with previous treatments of quantum indistinguishability.
In particular, the connections to the projective module approach developed in
\cite{Papadopoulos2004,Reyes2006,Reyes-Papadopoulos2009,Paschke2001}, as well as to the Berry-Robbins
approach, are discussed. Based on our results, a general definition of ``spin observables'' in the context of
quantum indistinguishability will be proposed. We finish with some conclusions in section
\ref{sec:conclusion}.

Let us remark that a detailed construction of angular momentum operators for a magnetic monopole and for a
system of indistinguishable particles (as in the present paper) will appear in the Villa de Leyva proceedings
volume~\cite{Benavides2009}. The explicit calculations performed in \cite{Benavides2009} using the Hopf
bundle provide an additional motivation for the concrete realization of the projective plane as a homogeneous
space that we have chosen here.
\section{Remarks on Canonical Commutation Relations}
\label{sec:ccr} For simplicity, we will start by considering quantum mechanics in one spatial dimension. We
want to recall some crucial facts that link the geometry of the classical phase space to the form of the
canonical commutation relations (CCR). In one spatial dimension, the CCR are given by
\begin{equation}
\label{eq:CCR} \left[\hat q, \hat p \right]= i\hbar,  \;\;\left[\hat q, \hat q \right]=0=\left[\hat p, \hat p
\right].
\end{equation}
According to the correspondence principle, quantum observables (as self-adjoint operators acting on  a
Hilbert space) are obtained by means of a map `` $\;\hat \;$ " for which the position variable $q$ is
promoted to a multiplication operator $\hat q$ and for which the  momentum $p$, the canonical conjugate of
$q$, is promoted to a differentiation operator $\hat p= -i\hbar d/dx$. More generally, Dirac's quantization
conditions require the replacement of  classical observables $f$ (that is, functions on phase space) by
self-adjoint operators $\hat{f}$ acting on a Hilbert space $\mathcal H$, in such a way that the Poisson
Bracket of two classical observables is mapped to the commutator of the corresponding self-adjoint operators.
Furthermore, one would like to have the property
\begin{equation}\label{eq:von-neumann}
\widehat{\phi(f)}=\phi(\hat f),
\end{equation}
for sufficiently well behaved real functions $\phi$, say, for polynomials (von Neumann's rule). As is well
known, such a quantization program is not implementable (Groenewold-Van Hove's theorem). The proof of that
theorem requires some additional technical assumptions, which we prefer to leave aside at this moment. For
our purposes, a simple example will make the point clear.

Suppose that a quantization map `` $\;\hat \;$ " having the above mentioned properties exists. Then we can
compare the result of quantizing the  function $f(q,p)=(pq)^2$ in two different ways, as explained below.
Assume that the von Neumann rule (\ref{eq:von-neumann}) holds for $\phi(t)=t^2$, and consider the identities
\begin{equation}
\label{eq:id1} pq=\frac{1}{2}\left((p+q)^2-p^2-q^2\right)
\end{equation}
and
\begin{equation}
\label{eq:id2} p^2q^2=\frac{1}{2}\left((p^2+q^2)^2-p^4-q^4\right).
\end{equation}
Applying the von Neumann rule to (\ref{eq:id1}) we obtain $\widehat{(pq)}=1/2(\hat p \hat q+\hat q\hat p)$.
Squaring this expression and applying the CCR and the von Neumann rule repeatedly, we then obtain:
\begin{equation}
(\widehat{pq})^2=\hat p^2\hat q^2 +2i\hbar \hat p \hat q -\frac{1}{4}\hbar^2.
\end{equation}
In contrast, an analogous procedure, starting from (\ref{eq:id2}), leads to
\begin{equation}\widehat{(p^2q^2)}=\hat p^2\hat q^2 +2 i\hbar \hat p\hat q -\hbar^2,
\end{equation}
showing clearly that there is a consistency problem. Mechanisms have nevertheless been devised in order to
circumvent this and similar difficulties. These have been available for a long time and are well known,
specially in the mathematical physics community. One of these, geometric
quantization~\cite{Souriau:69,Woodhouse:80}, is closely related to symplectic geometry. It allows one (in
certain cases), starting from a symplectic manifold $M$, to map homomorphically some subalgebra of the
Poisson algebra $(C^\infty(M),\{\, ,\,\})$ to an algebra of operators acting on a suitably defined Hilbert
space. But there are topological obstructions to the existence of such a map and the crucial steps of
polarization and half-form corrections often obscure the physical aspects of the problem.

Now, one could argue that quantization methods have little to do with the problem of quantum
indistinguishability, given that we are not only interested in the spin zero case, and that, as is often
said, ``there is no classical model  for spin''. Although it is possible to envisage classical models (i.e.
symplectic manifolds) for which spin arises as a consequence of a quantization procedure, this is not our
main point of concern, regarding the relevance of quantization. Our point is rather that, just because of the
fact that we are dealing with a \emph{classical configuration space}, the very definition of the self-adjoint
operators related to the symmetries of the configuration space depend crucially on the geometry of the
configuration space. And this is precisely where the experience gained from the development of the different
quantization methods might prove useful.

Quite recently, in a series of interesting papers, H.A. Kastrup~\cite{Kastrup2003,Kastrup2006,Kastrup2006a}
has been insisting on this point. He has pointed out that the careful consideration of these matters  for the
formulation of quantum mechanics on spaces like e.g.,  a circle, might have important consequences for
diverse problems  in quantum optics, in treatments of the Casimir effect, in relation to the cosmological
constant, etc.. The approach followed by Kastrup is to a great extent based  on the quantization method
explained in the inspiring Les Houches lecture notes by Isham~\cite{Isham1984}. This method, in turn, has
many features in common with geometric quantization and also uses  techniques developed by Mackey
\cite{Mackey1968} and Kirillov \cite{Kirillov1976}.

Returning to CCR, let us consider the following interesting example. Instead of $\mathbb R$, we take as
configuration space a half-line, the set $\mathbb R_+$ of positive real numbers. Let us assume that we can
take, as Hilbert space, the space $\mathcal H=(\mathbb R_+, dx)$. Now, let us assume for a moment that the
momentum operator $\hat p=-i\hbar d/dx$ is a well defined, self-adjoint operator. Then, since $\hat p$ is the
infinitesimal generator of translations, we can construct the unitary operator $U(a)=e^{-ia\hat p}$. This
operator would have the effect, on wave functions, of translating them by a distance $\hbar a$:
\begin{equation}
(U(a)\psi)(x)=\psi(x-\hbar a).
\end{equation}
But were this true, we could always choose $a$ in such a way that the support of $U(a)\psi$ ends up lying
outside $\mathbb R_+$. Thus, the CCR in the form (\ref{eq:CCR}) cannot hold in this space. The reason for the
breakdown of the CCR is that the vector field $\partial/\partial x$ on this space is not \emph{complete}.
Thus, the momentum operator $\hat p$ is not well defined. Another example, that has been much more discussed
 in the literature, is the circle. If we take as configuration space the circle $S^1$, the CCR breaks down
 again. The reason, in this case, lies in the fact that $S^1$ does not admit a global system of coordinates:
 At least two coordinate charts are needed. This fact has been known for a long time, the example being
  closely related, for instance,  to the Aharonov-Bohm effect. In this case, it is the \emph{topology} of the
  configuration space that determines the correct substitute for the CCR. Additionally, in this case there are many
  different (i.e. unitarily inequivalent) Hilbert space \emph{representations} of the new commutation relations, in contrast
  to the case of $\mathbb R^n$, where the Stone-von Neumann theorem guarantees the uniqueness, up to unitary equivalence, of the standard form
  of the CCR (\ref{eq:CCR}). Quite surprisingly, it is
  only until recently that these ideas have started to really draw  attention regarding concrete applications
  as,  for instance, in problems related to coherent states.

The particular emphasis of Isham's method on -first of all- finding the appropriate commutation relations,
given a classical configuration space, is very useful when studying this type of problems. It turns out that
even the CCR for quantum mechanics on $\mathbb R$ have a deep geometric origin. To explain this, let us
consider, following Isham, the unitary operators
\begin{equation}
\label{eq:2.3} U(a):=e^{-ia\hat{p}}, \,\;\;V(b):=e^{-ib\hat{q}}.
\end{equation}
Then, it is easy to check that the following transformation rules for the position and momentum operators
hold:
\begin{eqnarray}
\label{eq:2.4}
U(a)\hat{q}U(a)^{-1} & = & \hat{q}-\hbar a,\nonumber\\
V(b)\hat{p}V(b)^{-1} & = & \hat{p}+\hbar b.
\end{eqnarray}
Taking into account the correspondence principle, it is then natural to consider the following  action of the
additive group $\mathbb R\times \mathbb R$ on the phase space $T^*\mathbb R$:
\begin{eqnarray}
\label{eq:RnxRn}
(\mathbb{R}\times\mathbb{R})\times T^*\mathbb{R} & \longrightarrow & \;\;\;\;\;T^*\mathbb{R}\nonumber\\
(\,(a,b),(q,p)\,)\;\;& \longmapsto &(q-a,p+b).
\end{eqnarray}
A natural question then arises:

 What is the relation between the (additive) group
$\mathcal{G}=(\mathbb{R}\times\mathbb{R}, \,+\,)$ and the CCR in the form  $[\hat q,\hat p]=i\hbar$ which, in
the end, were the relations leading us to (\ref{eq:RnxRn})?

As explained at length in \cite{Isham1984}, the answer is given by a general formulation of the quantization
problem for a classical configuration space. According to Isham's method, the fundamental structure behind
the CCR is a so-called \emph{canonical group} $\mathcal C$. This group may arise directly from the geometry
of the configuration space, or more indirectly, arising as the solution to  an \emph{obstruction} problem,
related  to the construction of a quantization map. In both cases, there is a  group $\mathcal G$ of
transformations of the phase space, from where all other structures are derived\footnote{The groups $\mathcal
G$ and $\mathcal C$ are closely related and do coincide in many cases.}. In the example of $\mathbb R$, the
group $\mathcal G$ is $\mathcal{G}:=(\mathbb{R}\times\mathbb{R}, \,+\,)$, acting on phase space as indicated
above, and the group $\mathcal C$ is the \emph{Heisenberg group}. In the general case of $\mathbb R^n$, the
latter  is the group with underlying set $\mathbb R^n\times \mathbb R^n\times \mathbb R$ and product defined
by
\begin{equation}
(\vec a_1,\vec b_1, r_1)\cdot (\vec a_2,\vec b_2, r_2):=\left(\vec a_1+\vec a_2,\vec b_1+\vec b_2,
r_1+r_2+\frac{1}{2}(\vec b_1\cdot \vec a_2- \vec b_2\cdot \vec a_1)\right).
\end{equation}
The unitary representations of this group (returning to the case of $\mathbb R$) are given by operators of
the form
\begin{equation}\label{eq:U-Heisenberg}
\mathcal U(a,b,r)=U(a)V(b)e^{i\mu(r+ab/2)},
\end{equation}
with $U(a)$ and $V(b)$ operators satisfying the (Weyl) relations
\begin{eqnarray}
V(b_1)V(b_2) & =& V(b_1+b_2),\nonumber\\
U(a_1)U(a_2) &= & U(a_1+a_2),\\
U(a)V(b) &=& V(b)U(a)e^{-i\mu ab}.\nonumber
\end{eqnarray}
As is well known, these are precisely the commutation relations satisfied by the operators defined in
(\ref{eq:2.3}). The unitary representation $\mathcal U$ allows us to obtain a representation $\rho$ of the
Lie algebra of the canonical group $\mathcal C$. In this case, the Lie algebra of $\mathcal C$ is the
\emph{Heisenberg algebra}, which is given, for general $n$, by $(\mathbb R^n\oplus\mathbb R^n\oplus \mathbb
R, [\;,\;])$, with Lie bracket
\begin{equation}
[(\vec a_1,\vec b_1, c_1), (\vec a_2,\vec b_2, c_2)]:=\left(\vec 0,\vec 0, \vec b_1\cdot \vec a_2- \vec
b_2\cdot \vec a_1\right).
\end{equation}
As can be easily checked, the representation $\rho$ of this algebra obtained from $\mathcal U$, gives
precisely the CCR (\ref{eq:CCR}), with $\hat p= i \rho(1,0,0)$, $\hat q= i\rho(0,1,0)$ and $i\rho
(0,0,1)=\mu\equiv\hbar$. The answer to the question posed above is,  then: The Lie algebra of the canonical
group $\mathcal C$ is a \emph{central extension}, by $\mathbb R$,  of the Lie algebra of $\mathcal
G=(\mathbb{R}\times\mathbb{R}, \,+\,)$. Therefore, in this context, Planck's constant is seen to arise as the
``central charge'' of the extension. The deep geometrical meaning of these structures becomes clear after
examining different examples of configuration spaces, like the ones discussed above, and for which no \emph{a
priori} given position or momentum operators are given. Hence, in those cases, Isham's approach turns out to
provide a mathematically rigorous and physically profound method to find the appropriate commutation
relations. For instance, when the configuration space is the circle $S^1$, so that the phase space is
$T^*S^1\cong S^1\times \bb R$, the canonical group will have a subgroup given by the group of rotations
$SO(2)$, acting through diffeomorphisms and, in addition, an additive subgroup ($\bb R^2,+$) related to the
functions $\cos\phi$ and $\sin\phi$. The combined action of these groups on the phase space $T^*S^1$ gives
rise to a semi-direct product $\bb R^2 \rtimes SO(2)$ or, more generally, to the group $\bb R^2 \rtimes
\widetilde{SO}(2)$. The topological quantum effects related to this system (i.e. ``$\theta$-states'') can be
directly related to the unitary, irreducible representations of the \emph{canonical} group $\bb R^2 \rtimes
\widetilde{SO}(2)$. Similarly, Dirac's quantization condition for the charge of a magnetic monopole can be
directly linked to the unitary, irreducible representations of the group $\bb R^3\rtimes SU(2)$, the
canonical group of the phase space $T^*S^2$. In all these examples, the self-adjoint generators of the
representations will furnish an algebra whose commutation relations can be regarded as the ``canonical''
commutation relations appropriate for the given system.

In the next section, we will briefly describe Isham's method, and in section \ref{sec:RP2} we will use this
method  in order to construct the infinitesimal generators of rotations for a system of two identical
particles.

\section{Canonical Group Quantization}
\label{sec:isham} In this section we briefly summarize  the method of Canonical Group Quantization, as
developed by Isham in  ~\cite{Isham1984}.  The interested reader is encouraged to consult this reference,
where many examples are worked out explicitly and lots of background motivation is given.
\subsection{The canonical group}
Roughly speaking, the general idea of this method suggests that the quantization of a symplectic manifold $M$
is made possible, first, by finding a Lie algebra which is related (in a way to be explained) to a group
$\mathcal{G}$ of symplectic transformations of $M$ and, second, by the assumption that this Lie algebra
generates, in some sense, the set of classical observables. In other words, the idea of the quantization
scheme is to map isomorphically the Lie algebra $\mathcal{L(G)}$ onto some Lie subalgebra of
$(C^\infty(M,\mathbb{R}),\{\,,\,\})$. This map, called $P$, allows one to define a quantization map by fixing
a representation $U$ of the group and assigning to each function lying in the  image of $P$ the self-adjoint
generator obtained from $U$ by means of $P^{-1}$.

The program can be summarized by the following diagram:

\label{sec:quantization}
\begin{equation}
\label{diag:1} \hspace{-2cm}\xymatrix{& &{0}\ar[r]&{\mathbb{R}}\ar[r]&{C^\infty(M,\mathbb{R})}\ar[r]^{j}
&\mbox{HamVF}(M)\ar[r]&{0} .\\
& & & & & \mathcal{L}(\mathcal{G})\ar[u]_{\gamma}\ar@{-->}[ul]^{P}&}
\end{equation}
The meaning of the different terms appearing in (\ref{diag:1}) is the following. $M$ is a symplectic manifold
(phase space). In many cases, and particularly in the example we consider in this paper, this phase space
 is a cotangent bundle, $M=T^* Q$, with  $Q$ a configuration space.  The map $j$ assigns to each
 function $f$ on phase space (the negative of) its Hamiltonian vector field. Following the notation in
 \cite{Isham1984}, we shall write this action in the following way: $j(f)=-\xi_f$.
$\mathcal{G}$ is a Lie group, acting by symplectic transformations on $M$. $\mathcal{L}(\mathcal{G})$ denotes
the Lie algebra of $\mathcal{G}$. The map $\gamma: \mathcal{L}(\mathcal{G}) \rightarrow \mbox{HamVF}(M)$ is
the Lie algebra  homomorphism induced by the $\mathcal{G}$-action. The first row of the diagram represents a
short exact sequence, because the kernel of the map $j$ is the set of constant functions on phase space. The
motivation for studying that diagram comes from the fact that Hamiltonian vector fields generate local
one-parameter groups of symplectic transformations. Hence, if we are given a Lie sub-algebra of
$C^\infty(M,\mathbb{R})$, the corresponding Hamiltonian vector fields can, in principle, produce a group of
symplectic transformations (for this to be possible, the vector fields must be complete). In this way, a
relation between functions in the given Lie sub-algebra and elements in the Lie algebra of the group of
symplectic transformations can be obtained. The idea expressed in (\ref{diag:1}) is based on the possibility
of reversing the procedure, i.e., if we start with a group $\mathcal G$ of symplectic transformations, it is
possible to associate a locally Hamiltonian vector field to each element in the Lie algebra of the group.
This mapping is denoted $\gamma$ in (\ref{diag:1}). Now, if the vector fields are Hamiltonian, then one can
try to find a kind of ``inverse'' for $j$, i.e., to find a Lie algebra homomorphism $P:
\mathcal{L}(\mathcal{G})\rightarrow C^\infty(M,\mathbb{R})$ in such a way that
\begin{equation}
\label{eq:jPgamma} j\circ P= \gamma
\end{equation}
holds.

Summarizing, the first step of Isham's scheme consists in finding a Lie group $\mathcal{G}$ acting on $M$ by
symplectic transformations and then trying to associate its Lie algebra  with functions on $M$. More
precisely, one looks for a Lie algebra homomorphism
\begin{eqnarray}
P:  \mathcal{L}(\mathcal{G}) &\longrightarrow& C^\infty (M,\mathbb{R}) \nonumber \\
A &\longmapsto& P(A)
\end{eqnarray}
satisfying (\ref{eq:jPgamma}). In order for the $\mathcal{G}$-action on $M$ to be suitable for the
quantization procedure to work, the following requirements must be met (cf.~\cite{Isham1984}):
\begin{itemize}
\item The vector fields induced through $\gamma$ by the $\mathcal G$-action  must be globally Hamiltonian. This is
 required if the relation
(\ref{eq:jPgamma}) is to make sense and will be the case if $H^1(M,\bb R)=0$, or if $\mathcal{G}$  is
semi-simple.
\item The action must be almost effective (this implies that $\gamma$ will be injective) and transitive.
\end{itemize}
 If this task is accomplished, we will
be able to define a quantization map by using a unitary representation $U$ of the group and assigning to each
smooth function, in the image of $P$, the self-adjoint generator obtained from $U$.  To achieve the
correspondence between classical observables and the Lie algebra $\mathcal{L}(\mathcal{G})$, $P$ must be
linear and also a Lie algebra homomorphism. In other words, $P$ must satisfy, in addition to
(\ref{eq:jPgamma}),
\begin{equation}
\label{eq:2.5} \left\{P(A), P(B)\right\} = P(\left[A,B\right]),
\end{equation}
for all $A$ and $B$ in $\mathcal{L(G)}$. In general, the existence of a map $P$ with the desired properties
is not something obvious (cf. \cite{Isham1984}). There might be algebraic obstructions to the existence of
$P$, arising from the quite stringent requirement (\ref{eq:2.5}). In those cases where the obstruction cannot
be removed by a redefinition of the map $P$, a central extension of $\mathcal{L}(\mathcal{G})$ by $\bb R$ can
be used to construct the quantization map. In the example $Q=\mathbb R$, discussed in the previous section,
the obstruction cannot be removed and  this is the reason that forces one to consider the central extension
of the Lie algebra of $\mathcal G=(\mathbb R\times\mathbb R,+)$. The \emph{canonical group} $\mathcal{C}$
will be either $\mathcal{G}$ (in case the obstruction can be made to vanish) or otherwise it will be the
unique simply connected group the Lie algebra of which is the central extension of
$\mathcal{L}(\mathcal{G})$. The next step of the quantization method (briefly discussed in the next
subsection) consists in finding all irreducible, unitary representations of the canonical group.

An important question regarding the canonical group is whether, given a symplectic manifold $M$, we can find
a Lie group with the required properties and if so, how can we choose among the possible, candidate groups.
It turns out  that when the symplectic manifold is a cotangent bundle, $M=T^*Q$, there is a kind of
``universal'' solution to the problem of finding a canonical group. Indeed, the semi-direct product
$C^\infty(Q,\mathbb{R})/\mathbb{R} \rtimes \mbox{Diff}Q$  acts on $T^*Q$ by symplectic transformations. The
action $\rho$ is defined, for
 $[h]\in$ $C^\infty(Q,\mathbb{R})/ \mathbb{R}$, $\phi\in \mbox{Diff}Q$ and $l\in
T^*_qQ$, by:
\begin{equation}
\label{eq:2.action} \rho_{([h],\phi)}(l):= \phi^{-1*}(l)-(d h)_{\phi(q)}.
\end{equation}
Whereas the groups  $C^\infty(Q,\mathbb{R})/\mathbb{R}$ and  $\mbox{Diff}Q$ act separately on $T^*Q$ by
symplectic transformations in a natural (but not transitive) way, the combined action (\ref{eq:2.action})  is
a \emph{group action} if an only if the set $C^\infty(Q,\mathbb{R})/\mathbb{R}\times \mbox{Diff}Q$ is endowed
with the structure of a semi-direct product. This follows from the requirement $\rho_{g_1}\circ
\rho_{g_2}=\rho_{g_1g_2}$. Indeed, we have:
\begin{eqnarray}
 \rho_{([h_2],\phi_2)}\circ\rho_{([h_1],\phi_1)}(l)
&=& \rho_{([h_2],\phi_2)}(\phi_1^{-1*}(l)-(d h_1)_{\phi_1(q)})\nonumber\\
 &=&\phi_2^{-1*}(\phi_1^{-1*}(l)-(d h_1)_{\phi_1(q)})-
(dh_2)_{\phi_2\circ\phi_1(q)}\\
&=&((\phi_2\circ\phi_1)^{-1})^*(l)-d(\phi_2^{-1*}h_1+h_2)_{\phi_2\circ\phi_1(q)}\nonumber\\
&=&\rho_{([h_1\circ\phi_2^{-1}+h_2],\phi_2\circ\phi_1)}(l),\nonumber
\end{eqnarray}
so that $\rho$ is a group action if and only if the product in $C^\infty(Q,\mathbb{R})/\mathbb{R}\times
\mbox{Diff}Q$ is given by
\begin{equation}
([h_2],\phi_2)\cdot([h_1],\phi_1)=([h_2]+[h_1\comp\phi_2^{-1}],\phi_2\comp\phi_1 ).
\end{equation}
The crucial point is that  such a group action turns out to be transitive and effective so that, in this
case, the problem of finding a canonical group reduces to that of finding suitable finite dimensional
subgroups $W\leqslant C^\infty(Q,\mathbb{R})$ and $G\leqslant \mbox{Diff}Q$, such that the symplectic action
of $W\rtimes G$ on $T^*Q$ is still transitive.

\subsection{Representations of the canonical group for homogeneous spaces}
The next step of  Isham's scheme is to study the representations of the canonical group. In general, this is
not an obvious step, since the canonical group has non-trivial properties. However, we do not need to
consider the general case but only the class of groups that typically arise in physical systems. Actually, in
this work we only need to consider homogeneous spaces of the form $Q=G/H$ where $G$ and $H$ are Lie groups.
In this case, the procedure to follow is the following:
\begin{itemize}
\item[1.] Find a vector space $W$, and a linear action $R$ of $G$ on $W$, such that $G/H$ is
a $G$-orbit on $W$.
\item[2.] $T^*Q$ is then obtained by restriction of $T^*W\cong W\times W^*$.
\item[3.] There is  a left action of $\mathcal{G}:=W^*\rtimes G$ on $T^*W$, given by
\[l_{(\varphi',g)}(u,\varphi):= \left(R(g) u, R^*(g^{-1})\varphi -  \varphi'\right),\] where $R^*$ is the
linear action of $G$ on $W^*$ induced by duality. If we restrict the domain of every $\varphi\in W^*$ to the
$G$-orbit, we can regard $\varphi$ as a smooth function $f^\varphi: Q\rightarrow \bb R$, given by
$f^\varphi(u):=\varphi(u)$, ($u\in Q\hookrightarrow W$). Therefore, we can regard $\mathcal{G}$ as a finite
dimensional subgroup of $C^\infty(Q,\mathbb{R})/\mathbb{R}\times \mbox{Diff}Q$ and consider it as a
legitimate candidate for the canonical group.
\item[4.] The map $P$ from (\ref{diag:1}) is given, in this case, by
\begin{eqnarray}
\label{eq:2.8}
P: \mathcal{L}(W^*\rtimes G)& \longrightarrow & C^\infty(Q^*W,\mathbb{R})\nonumber\\
 \tilde A=(\varphi,A) \;\;\;\;&\longmapsto & P(\tilde A):(u,\psi)\mapsto \psi\left(R(A)u\right)+\varphi(u).
\end{eqnarray}
\item[5.] It can be shown (cf.~\cite{Isham1984}), that $\mathcal G$ satisfies all the properties required by
the scheme: The action is symplectic (by construction) and it is also effective and transitive. Moreover, the
map $P$ is a Lie algebra homomorphism. Hence, there is no obstruction and the canonical group can be chosen
to be $\mathcal G$.
\end{itemize}
 The unitary, irreducible representations of $\mathcal G$ can be constructed using Mackey's theory of induced
representations.
 The representation space will be the  space of square-integrable  sections of a hermitian vector bundle $E$ over $Q=G/H$,
 associated to the principal bundle $G\rightarrow G/H$. This requires the use of an irreducible unitary
 representation of $H$ and the existence of a $G$-quasi-invariant measure $\mu$ on $Q$.
  The operators representing $G$ are then constructed using a lift $l^\uparrow$ of the $G$-action $l$ on $Q$ to
   the corresponding associated vector bundle $E$.
This lift will provide $E$ with the structure of a $G$-vector bundle, i.e., the lift $l^\uparrow$ gives a
$G$-action on $E$ which is linear on the fibers and such that the following diagram commutes ($g\in G$):
\begin{equation}
\xymatrix{
  & & & {E}\ar[r]^{l^{\uparrow}_g}\ar[d]^{\pi}&  E\ar[d]_{\pi} & & &\\
& & &    Q \ar[r]^{l_g}                       &   Q            & & \\
    }
\end{equation}
In the present case, the lift is the one  naturally  induced by the right action of $G$ on the principal
bundle (cf.~\cite{Isham1984}). Explicitly, we have, for an equivalence class $[(p,v)]\in E$ (recall that $E$
is associated to $G\rightarrow G/H$ through an irreducible representation of $H$):
\begin{equation}\label{eq:natural-lift}
l^{\uparrow}_g([(p,v)]):=[(gp,v)].
\end{equation}
 Once we have identified
the lift, we can define the unitary operator $\mathcal U(\varphi,g)$ (this operator is to be compared to the
one defined in (\ref{eq:U-Heisenberg}) ) through its action on square-integrable sections $\Psi\in\Gamma(E)$:
\begin{equation}
\label{repre:sec} (\mathcal U(\varphi,g)\Psi)(x) :=e^{-i\varphi(x)}\sqrt{\frac{d\mu_g}{d\mu}(x)}\;
l^{\uparrow}_g \Psi(g^{-1}\cdot x),
\end{equation}
where $d\mu_g/d\mu$ is the Radon-Nikodym derivative of $\mu_g$ with respect to $\mu$.

In the present paper we are only interested in the subgroup $G$ of the canonical group, as it this this group
that will give rise to the topological quantum effects we are seeking. Let us therefore define the unitary
operator
\begin{equation}
U(g):=\mathcal U(0,g).
\end{equation}
 Now, if we consider a curve $g_t$ on $G$, corresponding to a given infinitesimal generator $J$ of the
group, we can obtain the action of the infinitesimal generators on sections through differentiation with
respect to $t$:
\begin{equation}
\label{repre:infinit} J\,\Psi(x):= \frac{d}{dt}\bigg|_{t=0}(U(g)\Psi)(x).
\end{equation}
This is the key formula for the calculations in the following sections.
\begin{remark}
Here we have considered the quantization of a configuration space  which is a homogeneous space $Q=G/H$.
Hence, there is a principal fibre bundle from which important information is extracted when constructing the
corresponding quantum theory:
\begin{equation}
H\hookrightarrow G\rightarrow G/H=Q.
\end{equation}
This point of view is particularly useful if one is interested in the quantum operators corresponding to the
self-adjoint generators of the group $G$. There is, though, another possible route, that puts more emphasis
on the role of the fundamental group of $Q$. Let $\widetilde Q$ denote the universal cover of $Q$. Then there
is a fibration
\begin{equation}
\pi_1(Q)\hookrightarrow \widetilde Q \rightarrow Q,
\end{equation}
for which there are certain \emph{lifting theorems} available. These theorems can be used to construct a lift
of the $G$-action on $Q$ to an associated vector bundle $\widetilde Q\times_\rho \bb C^k$, where $\rho$ is an
irreducible representation of $\pi_1(Q)$ on $\bb C^k$. When the subgroup $H$ is disconnected, interesting
quantum effects will appear (``$\theta$-states'') and, since they are closely related to the non-triviality
of $\pi_1(Q)$, the second approach provides an alternative and clear way to describe these effects.
\end{remark}
\section{Quantization of the Configuration Space of 2 Indistinguishable Particles}\label{sec:RP2}
\subsection{The configuration space}
The configuration space for a system of $N$ identical particles is not the Cartesian product $\bb{R}^{3N}$,
but the space obtained by identifying points in $\Bbb{R}^{3N}$ representing the same physical configuration.
Therefore, the configuration space can be written as
     \begin{equation}
     \label{eq:conf-space}
Q_{N} = \widetilde{Q}_{N}/S_{N},
\end{equation}
where
\begin{equation}
\widetilde{Q}_{N} = \left\{(r_{1},...,r_{N})\in\Bbb{R}^{3N}\,|\,r_{i}\neq r_{j} \,  \mbox{ whenever }\, i\neq
j\right\},
\end{equation}
and where $S_{N}$ is the permutation group. The non-coincidence condition $i\neq j$ is included in the
definition in order to make $\widetilde{Q}_{N}$ a manifold and to avoid the coincidence of two particles.
\begin{remark}
The non-coincidence condition is a topologically non-trivial assumption. It is because of this fact that the
fundamental group of $Q_N$ is isomorphic to $S_N$. It is known that the only two possible scalar
quantizations of $Q_N$ are those that give rise to Fermi or Bose statistics, and this in turn follows from
the fact that $\mbox{Hom}(\pi_1(Q_N),U(1))$ has exactly two elements. For an interesting discussion on the
physical status of this assumption, the reader is encouraged to consult \cite{Sorkin1992}.
\end{remark}
 In the
case of two identical particles, we can consider a transformation to center of mass and relative coordinates:
\begin{equation}
(\vec r_1,\vec r_2)\longmapsto \big(\vec R=\frac{1}{2}(\vec r_1+\vec r_2), \vec r =\vec r_1-\vec r_2\big).
\end{equation}
This transformation gives rise to a diffeomorphism $\bb R^3\times \bb R^3\cong \bb R^3_{\mbox{\small
cm}}\times \bb R^3_{\mbox{\small rel}}$. The non-coincidence condition does not play any role for the center
of mass position vector, whereas for the relative position vector $\vec r$ it implies $\vec r\neq 0$. For
this reason, when imposing the non-coincidence condition, we see that the space $\widetilde{Q}_2$ is
diffeomorphic to $\bb R^3\times S^2\times \bb R_+$. The factor $\bb R_+$ accounts for the relative distance
between the particles, i.e. we must have: $r=\parallel\vec r\parallel>0$. The action of the permutation group
on $\widetilde Q_2$ affects only the $S^2$ factor. Hence, when taking the quotient, we obtain:
 \begin{equation}
 Q_2 \cong \Bbb{R}^3 \times  \Bbb{R}P^2\times \bb R_+,
 \end{equation}
 where $\Bbb{R}P^2$ is the two dimensional projective space, appearing here in the form $S^2/\bb Z_2$.
\begin{remark}
As remarked above, it is the fundamental group of the configuration space that is responsible for the
topological quantum effects we are interested in. Although one could proceed to study the quantization of the
whole configuration space $Q_2$, it will be more convenient to isolate only the part of the canonical group
that is of direct relevance for our discussion. Nevertheless, it is important to emphasize that the
implementation of Isham's method to the quantization of $Q_2$ requires the use of some non-trivial additional
mathematical results (see below).
\end{remark}
Since our aim is to construct the infinitesimal generators of
 rotations for this problem, it will be  convenient to describe the
 configuration space as a homogeneous space for the group $SU(2)$. Since $\bb R P^2$ is the quotient of a sphere
 with respect to the antipodal map, we can use the fact that $S^2\cong SU(2)/U(1)$ to obtain a diffeomorphism
 of the form $SU(2)/H$, where $H$ is a \emph{disconnected} group. The details of this construction are given
 in the appendix at the end of the paper, and the result is that if we consider the following subgroup of
 $SU(2)$,
 \begin{equation}
H: = \left\{ \left( \begin{array}{ccc}
\lambda & 0 \\
0 & \bar{\lambda}
\end{array} \right),
\left( \begin{array}{ccc}
0 & \bar{\lambda} \\
-\lambda & 0
\end{array} \right) | \, \, \, |\lambda|^2 = 1  \right\},
\end{equation}
we obtain $\bb R P^2 \cong SU(2)/H$.  We denote the elements of $SU(2)$ by tuples $(z_0,z_1)$ that represent
matrices of the form
\begin{equation}
\left(
\begin{array}{cc}
z_0 & \bar z_1\\
-z_1 & \bar z_0
\end{array}
\right).
\end{equation}
Let us consider $U(1)$ as the subgroup of $SU(2)$ consisting of all diagonal matrices of the form
$\mbox{diag}(\lambda,\bar\lambda)$, with $|\lambda|=1$. A right action of $U(1)$ on $SU(2)$ is then given by
\begin{equation}
\left(
\begin{array}{cc}
z_0 & \bar z_1\\
-z_1 & \bar z_0
\end{array}
\right) \longmapsto \left(
\begin{array}{cc}
z_0 & \bar z_1\\
-z_1 & \bar z_0
\end{array}
\right) \left(
\begin{array}{cc}
\lambda & 0\\
0 & \bar \lambda
\end{array}
\right)=\left(
\begin{array}{cc}
\lambda z_0 & (\overline{\lambda z_1})\\
-(\lambda z_1) & (\overline{\lambda z_0})
\end{array}
\right),
\end{equation}
that can be equivalently expressed as $(z_0,z_1)\longmapsto (z_0,z_1)\cdot \lambda= (\lambda z_0,\lambda
z_1)$. Accordingly, points in the quotient space $SU(2)/U(1) \cong S^2$ will be denoted by $[z_0 : z_1]$.
This allows us to express  points in $\mathbb RP^2$ as equivalence clases in  $S^2/\mathbb{Z}_2$, of the form
$[[z_0 : z_1]]$.
\subsection{The canonical group}
For simplicity, we will only consider that part of the configuration space that corresponds to the relative
motion of the particles. Therefore, let us consider the manifold
\begin{equation}
\bb R P^2\times \bb R_+.
\end{equation}
The symplectic space we must consider is therefore $T^*(\bb R P^2\times \bb R_+)$. The symplectic 2-form is
therefore obtained, in a canonical way, from Liouville's 1-form, in the standard way.  We can use the fact
that the projective space can be written as a quotient of $SU(2)$ in order to construct the canonical group.
The $\bb R_+$ factor will contribute to the canonical group with certain ``dilation'' factors that pose no
problem. But, if we want to apply the techniques discussed on section \ref{sec:isham}, (especially the
techniques that apply for homogeneous spaces) we are confronted with the problem of finding a vector space
$W$ carrying a representation of $SU(2)$ in such a way that $\bb R P^2$ can be realized as an $SU(2)$-orbit
on $W$. Fortunately, there a theorem, due to Palais~\cite{Palais1960} and Mostow~\cite{Mostow1957}, which
guarantees that this can be done. The idea is to regard the space $C(\bb R P^2)$ as a representation space
for $SU(2)$. On then shows  that it is possible to find  an embedding $F:\mathbb R P^2\rightarrow \bb R^k$,
$F(x)=(f_1(x),\ldots,f_k(x))$, with the property that the vector spaces generated by the transformed
functions $g\cdot f_j$, with $g\in SU(2)$, are all finite dimensional. The dual of the direct sum of this
vector spaces carries, by construction, a linear action of $SU(2)$. It is finite dimensional and (this is the
key part) there is an $SU(2)$ orbit on it which is diffeomorphic to $\bb R P^2$. An explicit form for the
embedding is given (using homogeneous coordinates, with normalized entries), for instance, by
\begin{equation}
F([x:y:z]):=(yz,xz,xy,y^2-z^2).
\end{equation}
Since the component functions can be written as linear combinations of spherical harmonics, the embedding has
the required properties and gives rise to a linear action $R$ of $SU(2)$ on a vector space $W$ having an
orbit diffeomorphic to $\bb R P^2$. We can now use the linear action $R$ in order to construct the
semi-direct product
\begin{equation}
\mathcal G:= W^*\rtimes (SU(2)\times\bb R_+).
\end{equation}
Applying the techniques for homogeneous spaces described in section \ref{sec:isham}, we see that this group
can be used as a canonical group for the quantization of $\bb R P^2\times \bb R_+$. Using Mackey theory, one
finds that the unitary, irreducible representations of the canonical group relevant for the description of
spinless particles are defined on a (Hilbert) space of sections of a Hermitean line bundle over $\mathbb R
P^2\times \bb R_+$. The unitary operator corresponding to an element $(w,g,\lambda)\in W^*\rtimes
(SU(2)\times \bb R_+)$ of the canonical group, is the one acting on sections as follows:
\begin{equation}
\big (\mathcal U(w,g,\lambda)\big)\Psi([x],r):= \lambda^{3/2}e^{-i r
w([x])}l^\uparrow_g\Psi([g^{-1}x],\lambda r).
\end{equation}
The following remarks are in order:
\begin{itemize}
\item There are exactly two line bundles that can be chosen. One of them is the trivial one, and gives rise
to quantum particles obeying Bose statistics, and the other one is a non-trivial flat line bundle, that gives
 rise to quantum particles obeying Fermi statistics. The two bundles appear in this formalism as bundles
 associated to the principal bundle $H\hookrightarrow SU(2)\rightarrow SU(2)/H$ through unitary
 representations of $H$. For details, see below.
 \item Once a choice of line bundle has been made, we automatically obtain a lift $l^{\uparrow}$, using
(\ref{eq:natural-lift}).
\item Since we are mainly interested in the explicit form of the angular momentum operators, we will work
with the subgroup $SU(2)$ of the canonical group. For this reason, we define:
\begin{equation}
U(g):=\mathcal U(0,g,1).
\end{equation}
\end{itemize}
\subsection{Quantization of angular momentum generators}
 According to Isham's scheme, we must consider unitary representations of the group $H$. In one complex
dimension,
 we only have two possibilities. One given by the trivial representation and the other one given by
\begin{eqnarray}
\kappa:\hspace{1cm} H \hspace{0.7cm}&\longrightarrow& \mbox{Gl}(\Bbb{C}) \nonumber \\
\left( \begin{array}{ccc}
\lambda & 0 \\
0 & \bar{\lambda}
\end{array} \right) &\longmapsto& \;\;\;1,  \\
\left( \begin{array}{ccc}
0 & \bar{\lambda} \\
-\lambda & 0
\end{array} \right) &\longmapsto& -1 \nonumber.
\end{eqnarray}
The total space of the   line bundle $SU(2) \times_{\kappa} \Bbb{C}$ associated to the principal bundle
$SU(2)\rightarrow SU(2)/H$ is the space $\left\{[(g,v)]\, | \, g\in SU(2) \, {\rm and} \, v \in
\Bbb{C}\right\}$ of equivalence classes defined by the equivalence relation $(g,v)\sim (g
h,\kappa(h^{-1})v)$. The projection is given by
\begin{equation}
\pi_\kappa\left([(z_0,z_1),v]\right)=[[z_0:z_1]].
\end{equation}
 The action of the
rotation group $SU(2)$ on $\mathbb R P^2$ is given, for $g=(\alpha,\beta)\in SU(2)$ and $p=[[z_0:z_1]]\in
\mathbb R P^2$, by $ l_g(p)=[[ \alpha z_0 - \bar{\beta}z_1: \beta z_0 + \bar{\alpha} z_1]]$. According to the
discussion of the previous section, a lift is naturally induced by the principal bundle $SU(2)\rightarrow
SU(2)/H$. Explicitly, we have:
\begin{equation}
l^\uparrow_g\left([(z_0,z_1),v]\right)= \left[\left(g(z_0,z_1), v\right)\right].
\end{equation}
The action of the corresponding infinitesimal generators can now be explicitly computed, following the
prescriptions (\ref{repre:sec}) and (\ref{repre:infinit}).  For more details on this calculation, the reader
is referred to \cite{Benavides2009}.
\section{Interpretation of the Results and Comparison with Other Approaches}\label{sec:other}
\subsection{Explicit expressions for the infinitesimal generators}
Having the above remarks in mind, let us proceed to obtain explicit expressions for the infinitesimal
generators constructed in the previous section. We start by constructing an explicit isomorphism between the
bundle $SU(2)\times_\kappa \mathbb{C}$ and the nontrivial  line bundle $\mathcal{L}_-$ of $\mathbb{R}P^2$,
regarded  as a subbundle of the trivial bundle $\mathbb{R}P^2\times \mathbb{C}^3\rightarrow \mathbb{R}P^2$,
as described below.

Using the standard homogeneous coordinates, we cover $\mathbb{R}P^2\cong S^2/\mathbb Z_2$ with the following
charts ($\alpha=1,2,3$):
\begin{equation}
\label{cartasloc} U_{\alpha} =  \{[x] \in \mathbb{R}P^2\, |\; x_\alpha \neq 0\}.
\end{equation}
The  total space of  $\mathcal{L}_{-}$ is, as a set, given by:
\begin{equation}
\label{eq:4.1} \left\{\left(\left[x\right], \lambda\left|\phi(x)\right\rangle\right)\in \mathbb{R}P^2 \times
\Bbb{C}^3 \, | \, \lambda \in \Bbb{C} \, {\rm and} \, x \in \left[x\right]\right\},
\end{equation}
with
\begin{eqnarray}
 |\phi(-)\rangle : S^2 & \longrightarrow & \mathbb{C}^3 \nonumber\\
 x &\longmapsto &|\phi(x)\rangle
\end{eqnarray}
being \emph{any} map from $S^2$ to  $\mathbb{C}^3$ satisfying the following conditions:
\begin{itemize}
\item[({\it i})] It is smooth.
\item[({\it ii})] $|\phi(x)\rangle\neq  0$ for all $x\in S^2$.
\item[({\it iii})] $|\phi(-x)\rangle = -|\phi(x)\rangle$ for all $x\in S^2$.
\end{itemize}
The bundle projection is defined through
 $\pi\left(\left(\,\left[x\right], \lambda\left|\phi(x)\right\rangle\,\right)\right)=\left[x\right]$.
According to (\ref{eq:4.1}), an element in the total space of $\mathcal{L}_-$ is given by a tuple of the form
$([x],\lambda|\phi(x)\rangle)$. Alternatively, we can describe the bundle saying that the fiber over $[x]$ is
the subset
  $\lbrace[x]\rbrace\times V_{[x]}$ of $\mathbb{R}P^2\times \mathbb{C}^3$, where $V_{[x]}$ is the vector
  space generated by the vector $|\phi(x)\rangle\in \mathbb{C}^3$.
Local trivializations for $\mathcal{L}_-$ are given by ($\alpha=1,2,3$):
\begin{eqnarray}
\varphi_\alpha : \pi^{-1}(U_\alpha) &\longrightarrow& U_\alpha \times \Bbb{C} \nonumber \\
\left(\left[x\right], \lambda\left|\phi(x)\right\rangle\right) &\longmapsto& \left(\left[x\right], {\rm
sign}(x_\alpha)\lambda\right).
\end{eqnarray}
They give rise to the following transition functions:
\begin{eqnarray}
g_{\alpha\beta}: U_\alpha \cap U_\alpha & \longrightarrow & \mathbb{Z}_2\leqslant U(1)\nonumber\\
\left[\,x\right]&\longmapsto &g_{\alpha\beta}([x])= \mbox{sign}(x_\alpha x_\beta).
\end{eqnarray}

If  $g=(z_0,z_1)\in SU(2)$ and $v\in\mathbb{C}$, then $\pi_\kappa([(g,v)])=[[z_0:z_1]]$ is a point in
$SU(2)/H$. Let $x(g)$ denote the point in $S^2$ obtained from $g$ through the quotient map $SU(2)\rightarrow
SU(2)/U(1)$ and let $[x(g)]$ denote the corresponding equivalence class, with respect to the quotient map
$S^2\rightarrow S^2/\mathbb{Z}_2$. Then it is clear that $\pi_\kappa([(g,v)])=[x(g)]$, independently of the
chosen $g$. This fact allows us to construct the following isomorphism between the two bundles:
\begin{eqnarray}
\Phi:   SU(2) \times_{\kappa} \Bbb{C} &\longrightarrow& \;\;\;\;\;\mathcal{L}_{-} \nonumber\\
\left[(g, v)\right]\;\;\; &\longmapsto& \left(\left[x(g)\right], v\left|\phi(x(g))\right\rangle\right).
\end{eqnarray}
Using this isomorphism, we can ``transfer'' the lift to the bundle $\mathcal{L}_-$, as indicated in the
following diagram:
\begin{equation}
\xymatrix{
  & & & \mathcal{L}_{-} \ar @{-->}[rr]^{\tau_g}
 \ar[d]_{\Phi^{-1}}
 &
  & \mathcal{L}_{-}  & &\\
  & & & {SU(2) \times_{\kappa} \Bbb{C}}\ar[rr]^{l^{\uparrow}_g}
 \ar[d]^{\pi_\kappa}
 &
  & SU(2) \times_{\kappa} \Bbb{C} \ar[u]^{\Phi} \ar[d]_{\pi_\kappa} & & &\\
   & & &
\mathbb{R}P^2 \ar[rr]^{l_g}& &
\mathbb{R}P^2  & & \\
    }
\end{equation}
From $\tau_g = \Phi \circ l^{\uparrow}_g \circ \Phi^{-1}$ we get, for $g=(\alpha,\beta)$:
\begin{eqnarray}
\label{eq:lift} \tau_g\left(\left[x\right], \lambda \left|\phi(x)\right\rangle\right) &=&
(\Phi \circ l^{\uparrow}_g \circ \Phi^{-1}) \left(\left[x\right], \lambda \left|\phi(x)\right\rangle\right) \nonumber \\
&=& (\Phi \circ l^{\uparrow}_g ) \left[(\left(z_0, z_1\right), \lambda)\right] \nonumber \\
&=& \Phi \left[(\left(\alpha, \beta\right) \cdot \left(z_0, z_1\right), \lambda)\right] \nonumber \\
&=& \left(\left[x(z^{'}_0, z^{'}_1)\right], \lambda \left|\phi(x(z^{'}_0, z^{'}_1))\right\rangle\right),
\end{eqnarray}
where we have used the notation $z^{'}_0 = \alpha z_0 - \bar{\beta}z_1$ and $z^{'}_1 = \beta z_0 +
\bar{\alpha} z_1$. Here, $(z_0,z_1)$ is chosen in such a way that $x\equiv x(z_0,z_1)=[[z_0:z_1]]$.

Now, notice that a smooth section on $\mathcal{L}_-$ can always be written in the form $\Psi([x\,]) =
(\left[x\right], a(x) \left|\phi(x)\right\rangle)$, with $a:S^2\rightarrow \mathbb{C}$ a smooth
\emph{antisymmetric} function~\cite{Reyes2006}. Such a section transforms under the action of $SU(2)$ in the
following way:
\begin{eqnarray}
(U(g)\Psi)([x]) &:=& \tau_g(\Psi(g^{-1}\cdot [x])) = \tau_g(\left[g^{-1}\cdot x\right],
 a(g^{-1}\cdot x) \left|\phi(g^{-1}\cdot x)\right\rangle) \nonumber \\
&=& \left(\left[x\right], a(g^{-1}\cdot x) \left|\phi(x)\right\rangle\right).
\end{eqnarray}
From this we immediately see that the infinitesimal generators $J_i$ are given by
\begin{equation}
\label{eq:4.2} (J_i\Psi)([x]) = \left(\left[x\right], (L_i a)( x) \left|\phi(x)\right\rangle\right),
\end{equation}
where $L_i$ is the usual (orbital) angular momentum operator!

\subsection{Interpretation of the results in the light of the projective module approach}
In order to interpret this result, let us make reference to the approach proposed in \cite{Paschke2001} and
further developed in \cite{Papadopoulos2004,Reyes2006,Reyes-Papadopoulos2009}, where extensive use of the
 Serre-Swan equivalence of vector bundles and projective modules is made. Since the use of this equivalence
 is (perhaps) not generally known in the context of non-relativistic quantum mechanics, the interested reader
 is referred to \cite{Papadopoulos2004,Reyes2006,Reyes-Papadopoulos2009,Paschke2001} where many explicit
 calculations are carried out.
   In the present case,
 given a (normalized) map $|\phi(-)\rangle$ satisfying
  properties ({\it i})-({\it iii}), it is possible to show
   that the \emph{projector} $p:[x]\mapsto |\phi(x)\rangle\langle\phi(x)|$
gives rise to a finitely generated projective module $p(\mathcal{A}_+^3)$ over the algebra $\mathcal{A}_+$ of
complex, continuous even functions over the sphere.  It can be shown that this module is isomorphic to
 the $\mathcal A_+$-module  of sections on the bundle $\mathcal{L}_-$: $\Gamma(\mathcal L_-)$. Let
  us explain this in some detail.

Regard the two-sphere $S^2$ as the effective configuration space (space of normalized relative coordinates)
for a system of two point particles in three spatial dimensions, excluding coincidence points. Then, the
configuration space for the two particles, when regarded as \emph{indistinguishable} is the projective space
$\mathbb RP^2$. Using the fact that every even function on $S^2$ can be regarded as a function on $\mathbb
RP^2$ and vice versa, it is useful to consider the following decomposition of the space $C(S^2)$ of complex
continuous functions on the sphere:
\begin{equation}
C(S^2)=\mathcal A_+\oplus \mathcal A_-,
\end{equation}
where (as already mentioned) $\mathcal A_+$ denotes the subspace of \emph{even} functions and $\mathcal A_-$
denotes the subspace of \emph{odd} functions. Clearly, we must have the following \emph{algebra} isomorphism:
\begin{equation}
\mathcal A_+ \cong C(\mathbb RP^2),
\end{equation}
or, equivalently:
\begin{center}
\begin{tabular}{|p{11cm}|}
\hline \\
\emph{Even} functions on $S^2$ can be regarded as functions on $\mathbb RP^2$ or, equivalently, as
\emph{sections} on the trivial line bundle $\mathcal L_+$ over $\mathbb RP^2$.
\\
\\
 \hline
\end{tabular}
\end{center}
On the other hand, we have the following isomorphisms of finitely generated, projective $\mathcal
A_+$-modules:
\begin{equation}
\mathcal A_- \cong p(\mathcal A_+^3)\cong \Gamma(\mathcal L_-),
\end{equation}
or, in other words,
\begin{center}
\begin{tabular}{|p{11cm}|}
\hline \\
\emph{Odd} functions on $S^2$ can be regarded as \emph{sections} on the non-trivial line bundle $\mathcal
L_-$ over $\mathbb RP^2$.
\\
\\
 \hline
\end{tabular}
\end{center}
According to the quantization approach we have used in this paper,  the Hilbert space of a theory obtained
from a classical configuration space $Q$ will be given, in general, by (the completion of) some space of
square integrable section on a given bundle over $Q$. That means that, in our present example, the two
Hilbert spaces we obtain are given by (a suitable completion) of the section spaces $\Gamma(\mathcal L_+)$
and $\Gamma(\mathcal L_-)$ (i.e. the Fermi-Bose alternative). But each one of these ``pre'' Hilbert spaces is
-in every respect- isomorphic, and hence carries exactly the same information, as the module $\mathcal A_+$,
respectively  $\mathcal A_-$. Therefore, if we consider the transformation properties of a wave function that
is a section in $\mathcal L_\pm$, we must obtain an equivalent transformation rule in the corresponding space
$\mathcal A_\pm$. The construction of infinitesimal generators for rotations that we have just performed is
based on a rigorous quantization scheme. One is thus comforted  to find that the angular momentum operators
$J_i$ obtained this way, as operators acting on $\Gamma(\mathcal L_-)$, exactly match the usual angular
momentum operators $L_i$, acting as differential operators on (the restriction to $\mathcal A_-$ of)
$C(S^2)$.

To be more explicit, let us call $\Phi$ the $\mathcal A_+$-module isomorphism between $\mathcal A_-$ and
$\Gamma(L_-)$. With a fixed choice  of $|\phi\rangle$,  as described above,  it is given by
\begin{eqnarray}
\Phi: \mathcal A_- &\longrightarrow & \Gamma(\mathcal L_-)\nonumber\\
a& \longmapsto&  \Psi_a,
\end{eqnarray}
where $\Psi_a([x\,]):= a(x)|\phi(x)\rangle$. Notice that in spite of the fact that $x\mapsto| \phi(x)\rangle$
\emph{cannot} be regarded as a section of $\mathcal L_-$, the map $[x\,]\mapsto \Psi_a([x\,])$ is a well
defined section on $\mathcal L_-$. What our results indicate, then, is not only that the underlying  linear
spaces from which the corresponding  Hilbert spaces are supposed to be constructed are mathematically
equivalent, but that the corresponding generators of symmetries do also coincide. In fact, (\ref{eq:4.2}) can
now be rewritten as follows: $J_i(\Psi_a)\equiv \Psi_{L_i(a)}$ or, in a more suggestive way:
\begin{equation}
J_i\circ \Phi= \Phi \circ L_i.
\end{equation}
Let us note that similar intertwining relations for angular momentum operators have been shown by
Kuckert~\cite{Kuckert:04} to play an important role regarding the spin-statistics connection. It would
therefore be desirable to obtain a global version, along the lines explained in the present paper, of his
results. Recalling the remarks made in sections \ref{sec:ccr} and \ref{sec:quantization}, we are led to
believe that important information is being disregarded when studying the spin-statistics connection using
the configuration space (\ref{eq:conf-space}), namely, the implementation (by means of self-adjoint
operators) of the angular momentum operators corresponding to the generators constructed above may involve
some representation-theoretic and functional analytic aspects that could be of importance. An example
reinforcing our point of view is given by the careful consideration of the possible self-adjoint extensions
of the Laplacian on these configuration spaces (for two spatial dimensions), as carried out some time ago by
Bourdeau and Sorkin \cite{Sorkin1992}.
\subsection{How the observable ``spin'' should be defined}
As a motivation, let us  begin this subsection by considering the following elementary examples.\medskip\\
\underline{Spin zero particle in three spatial dimensions:}\medskip\\
 The wave function of such a particle is a complex
function $\psi:\bb R^3\rightarrow \bb C$. Since there are no internal degrees of freedom, the action of a
rotation $g\in SU(2)$ on the wave function is simply given by
\begin{equation}
g\cdot \psi(\vec r) =\psi(g^{-1}\vec r).
\end{equation}
Notice that, while the left hand side is being evaluated at the point $\vec r$ in space, the right hand side
is being evaluated at a \emph{different} point, namely, $g^{-1}\vec r$. Whereas in this case this
transformation rule makes prefect sense, if the configuration space were a topologically non trivial
manifold, such that the more general possibility of defining the wave function as a section on a non trivial
bundle would exist, the above rule would not make sense. What one needs to correct that rule in the general
case is a lift $l^\uparrow$ of the $SU(2)$ action to the bundle. As we have seen, in the case of two
identical spin zero particles there are two possibilities for the quantum theory. The one corresponding to
Fermi statistics (i.e. the one giving the wrong spin-statistics connection) is implemented on a Hilbert space
of sections on a non trivial line bundle. Hence we see that, even if the particle do not have spin, a lift to
the total space will be generally needed.

The next example is intended to show that the transformation properties for \emph{one} particle of spin 1/2
already include a lift, although the bundle in this case is a trivial bundle.\\
\\
\underline{Spin $1/2$ particle in three spatial dimensions:}
\medskip\\
 In this case the Hilbert space is given by
\begin{equation}
\mathcal H= L^2(\mathbb R^3)\otimes \mathbb C^2.
\end{equation}
The spin state space $\mathbb C^2$ has a basis $\{|+\rangle,|-\rangle\}$, with respect to which the wave
function can be written:
\begin{equation}
\psi= \psi_+ \otimes |+\rangle + \psi_-\otimes |-\rangle, \nonumber
\end{equation}
with $\psi_+$ and $\psi_-$ position dependent functions. The spin operators $S_{\pm},S_3$, proportional to
the Pauli matrices, furnish a representation of the Lie algebra
 $\mathfrak{su}(2)$, acting on $\mathbb C^2$:
\begin{eqnarray}
S_{\mp}|\pm\rangle & = & |\mp\rangle \nonumber\\
S_{\pm}|\pm\rangle & = & 0 \\
S_{3}|\pm\rangle & = &\pm\frac{1}{2} |\mp\rangle.\nonumber
\end{eqnarray}
These matrices, together with the angular momentum operators $L_i$,  give rise to the infinitesimal
generators of rotations, acting  on $\mathcal H$, by angular momentum addition:
\begin{equation}
 J_i=L_i\otimes \mbox{Id}_{\mathbb C^2} + \mbox{Id}_{L^2(\mathbb R^3)}\otimes S_i.
\end{equation}
By exponentiation of $L_i$ and $S_i$ we obtain representations of the rotation group $SU(2)$ on $L^2(\mathbb
R^3)$ and $\mathbb C^2$, respectively. Explicitly ($g\in SU(2)$), we have:
\begin{eqnarray}
(g\cdot \psi_{\pm})(\vec r)&=&\psi_{\pm}(g^{-1}\cdot \vec r)\nonumber \\
g\cdot |\pm\rangle &=& \mathcal{D}^{1/2}(g)|\pm\rangle \nonumber,
\end{eqnarray}
where $\mathcal{D}^{1/2}(g)$ are the usual rotation (Wigner) matrices.  Thus, the transformation of the
complete wave function under finite rotations is given by
 $\psi\rightarrow g\cdot\psi$, with $g\cdot \psi$ defined by:
\begin{equation}\label{eq:g.psi}
(g\cdot \psi)(\vec r):= \mathcal{D}^{(1/2)}(g)(\psi(g^{-1}\cdot \vec r )).
\end{equation}
In this case, the wave function can be regarded as a vector valued map $\psi:\mathbb R^3 \rightarrow \mathbb
C^2$ or, equivalently, as a \emph{section} $\sigma_\psi$ on the trivial vector bundle with  total space given
by $\mathbb R^3\times \mathbb C^2$. The section corresponding to $\psi$ is then given by
\begin{equation}
\sigma_\psi (\vec r):= (\vec r, \psi(\vec r)).
\end{equation}
Looking back at (\ref{repre:sec}), we see that by regarding the wave function as a section of the (trivial)
bundle $\mathbb R^3\times \mathbb C^2\rightarrow \mathbb R^3$, the transformation rule (\ref{eq:g.psi})  is
the one induced by the following lift:
\begin{eqnarray}
\label{eq:g-E}
l^{\uparrow}: SU(2)\times (\mathbb R^3\times \mathbb C^2) &\longrightarrow & (\mathbb R^3\times \mathbb C^2)\\
(g, (\vec r, v))&\longmapsto & l^{\uparrow}_g(\vec r,v):=(g\cdot \vec r,\, \mathcal{D}^{(1/2)}(g)v)\nonumber.
\end{eqnarray}
We thus recognize that this familiar case  also fits in the general scheme discussed in the previous
sections.  Moreover, this shows that, in order to define the transformation properties of a wave function
which is given by a section in a (possibly non trivial) vector bundle over the configuration space, all we
need  is to find a lift of the group action to the total space of the bundle. This can have interesting
effects. For example, in the case of a scalar\footnote{In the sense of having spin zero} particle $Q=S^2$,
the behavior under rotations of the wave function can have \emph{fermionic} character, according to the
(induced) representation of $SU(2)$ chosen. In that case, this is a consequence of the non triviality of the
bundle~\cite{Benavides2009}.

Motivated by the rigorous result obtained by application of Isham's method to the case of spin zero
particles, we thus propose:
\begin{center}
\begin{tabular}{|p{11cm}|}
\hline \\
\emph{The correct definition of spin operators for a system of $N$ indistinguishable particles must involve a
lift of the $SU(2)$-action on the configuration space $Q_N$ to the vector bundle where the wave functions are
defined.}
\\
\\
 \hline
\end{tabular}
\end{center}
In the next subsection we will show how this definition fits perfectly with the definition given by Berry and
Robbins in \cite{Berry1997}.
\subsection{Relation with the Berry-Robbins approach}
Let us now briefly comment on how our approach relates to the one of Berry and Robbins~\cite{Berry1997}. In
this approach, spin vectors are position dependent, giving rise to a \emph{transported spin basis}. The
transition from the fixed spin basis to the transported one is effected through the action of a position
dependent unitary operator $U(\vec r)$. Momentum as well as spin operators are also obtained from the
``fixed'' operators, and thus required to depend on position. In particular, the infinitesimal generators for
\emph{spin angular momentum} are given by
\begin{equation}
S_i(\vec r):= U(\vec r) S_i U^\dagger(\vec r).
\end{equation}
In the present paper we have only considered the spin zero case. Nevertheless, it becomes apparent that for
general values of spin, the appropriate definition of (spin) angular momentum operators should involve the
structure of a $SU(2)$-bundle.  What the $SU(2)$ lift that we have considered for the spin zero case
accomplishes is, for a given $g\in SU(2)$, to generate the transport of a vector in the fiber over a given
point $q$ to some other vector, lying in the fiber over the image point $l_g(q)$. It can be
shown~\cite{Reyes2006} that, for $g$ close to the identity of the group, parallel transport (with respect to
the canonically given flat  connection on the bundle) of the vector from the fiber over $q$ to the fiber over
$l_g(q)$, coincides with the action of the lift, $l^\uparrow_g$, on the given vector. This  can be recognized
in (\ref{eq:lift}) from the effect of the lift on elements of the total space of $\mathcal L_-$, because
$([x],\lambda|\phi(x)\rangle)$ gets mapped to $(l_g([x]),\lambda|\phi(g\cdot x)\rangle)$. Were we considering
spin degrees of freedom, we would have more ``transported vectors'' of the type $|\phi\rangle$. They would
play the role of the transported spin basis of Berry and Robbins. Then, the effect of a rotation would
include, apart from this ``parallel transport rotation'', an instrinsic rotation within each fiber. This is
precisely what the transported spin operators $S_i(\vec r)$ accomplish. In the construction of Berry and
Robbins, parallel transport is effected by the operator $U(\vec r)$. Thus, we can recast the action of
$SU(2)$ on wave functions in that construction as follows. Let us denote with $|j,m(\vec r)\rangle$ the
transported spin vectors, written in the total angular momentum basis. If $\eta^s$ is the vector bundle
(constructed as a sub-bundle of a trivial bundle of higher rank) over the sphere, whose fibers over $\vec r$
is the vector space spanned by all the vectors $|j,m(\vec r)\rangle$ ($\vec r$ being kept fixed), then we can
define the following \emph{lift} of the $SU(2)$ action on the sphere:
\begin{equation}l_g^\uparrow:
\left(\;r\;,\sum_{j,m}\lambda_{j,m}|j,m(r)\rangle \right)\longmapsto
 \left(\;g\cdot r\;, \sum_{j,m}\lambda_{j,m}U(g\cdot r)\mathcal{D}^j(g)U(r)^\dagger|j,m(r)\rangle
\right).
\end{equation}
Following the prescriptions indicated in (\ref{repre:sec}) and (\ref{repre:infinit}), it is then possible to
recover the spin operators $S_i(\vec r)$ from the above defined lift. This shows that construction of Berry
and Robbins also fits into the framework described in this paper.

 Finally, we would like to contribute to clarify a point that, to our opinion,
has been misinterpreted. Recently, Allen and Mondragon~\cite{Allen2003} have criticized the proposal made by
Peshkin in \cite{Peshkin2003} (for Peshkin's reply, see \cite{Peshkin2003b}). While we do agree that
Peshkin's construction contains a flaw in the argumentation, we do not share the opinion of those authors
(and others, like in \cite{Streater:homepage}) who (referring to recent work on spin-statistics, particularly
to \cite{Peshkin2003} and \cite{Berry1997}) claim that \emph{``..quantum mechanics is modified so as to force
a spin-statistics connection, but the resulting theory is quite different from standard physics''}. In this
paper, we have tried to motivate and to explain why we believe such claims are completely unsubstantiated.
From the remarks on the previous sections, it is completely clear that quantum mechanics can be consistently
formulated on a great variety  of configuration spaces. While these schemes have,  with time, grown to be
full-fledged mathematical theories and so, many interesting examples do not seem to have a direct physical
interpretation or application, schemes like Isham's one go to the very heart of quantum mechanics and provide
useful and physically sensible results and insights. In particular, the equivalence between $\mathcal A_\pm$
and $\Gamma(\mathcal L_\pm)$ has been extended in this paper to include the infinitesimal generators of
rotations. Thus, while it is true that taking the indistinguishability of particles into account already at
the level of the configuration space makes the analysis of the corresponding quantum theory more complicated,
due to the non-trivial topology of the configuration space, a detailed study of the respective quantum theory
shows that we can keep exactly the same textbook formalism of quantum mechanics, with the conceptual
difference that the Fermi-Bose alternative (at least for spin zero particles) and with it the symmetrization
postulate, do not need to be imposed ``by hand''.
\section{Conclusions}\label{sec:conclusion}
In this work we have considered the canonical group quantization of the projective space $\mathbb R P^2$, as
a model for the quantum theory of two non-relativistic spin zero identical particles. The approach has been
motivated by a discussion of the relevance of commutation relations for the formulation of a quantum theory
based on a classical configuration space. The main focus has been on the construction of the infinitesimal
generators of rotations. The operators obtained have been used to complete the equivalence already
established between  the spaces of wave functions on the sphere and spaces of sections on the projective
space. The connection of our formalism with different approaches has been established and discussed.
\appendix
\section{}
 The projective plane $\bb RP^2$ can be represented as the quotient $SO(3)/O(2)$, where
$O(2)$ is the group of orthogonal transformations. This group can be written as the union of two disjoint
sets: The matrices with determinant $1$ and the matrices with determinant $-1$. Then, in order to show that
the projective space is a homogeneous space, we write the orthogonal group $O(2)$ as a subgroup of $SO(3)$:
\begin{displaymath}
\left\{ \left( \begin{array}{ccc}
\cos\phi & -\sin\phi & 0 \\
\sin\phi & \cos\phi & 0 \\
0 & 0 & 1
\end{array} \right),
\left( \begin{array}{ccc}
-\cos\phi & \sin\phi & 0 \\
\sin\phi & \cos\phi & 0 \\
0 & 0 & -1
\end{array} \right) | \, \, \, 0\leq \phi\leq 2\pi  \right\} .
\end{displaymath}
Since this is a subgroup of $SO(3)$, by means of the ``spinor map'' (here we follow the notation from
\cite{Nab}), we can associate each matrix in $O(2)$ with a matrix in $SU(2)$. In fact, the homomorphism
\begin{eqnarray}
{\rm Spin}: SU(2) &\rightarrow& SO(3) \nonumber \\
u(\psi, \hat{n}) &\rightarrow& R(\psi, \hat{n}),
\end{eqnarray}
relates each matrix in $SU(2)$ to a matrix in $SO(3)$, where
\begin{eqnarray}
u(\psi, \hat{n}) &=& \cos(\psi/2) \mbox{Id} - i \sin(\psi/2)(x\sigma_x + y \sigma_y + z\sigma_z) \nonumber \\
&=& \left( \begin{array}{ccc}
\cos(\psi/2) - i\sin(\psi/2)z & -y \sin(\psi/2) - ix\sin(\psi/2) \\
y \sin(\psi/2) - ix\sin(\psi/2) & \cos(\psi/2) + i\sin(\psi/2)z \nonumber
\end{array} \right).
\end{eqnarray}
Every rotation can be parameterized by an axis of rotation $\hat{n} = (x, y, z)$, with $|x|^2 + |y|^2 + |z|^2
= 1$, and an angle of rotation about this axis, $\psi$. Then, we can write every matrix in the rotation group
$R(\psi, \hat{n})$ by
\begin{equation}
\label{RenN} R(\psi, \hat{n}) = e^{\psi N} = \mbox{Id} + (\sin \psi) N + (1 - \cos \psi) N^2 ,
\end{equation}
where $\mbox{Id}$ is the identity matrix and
\begin{displaymath}
N = \left( \begin{array}{ccc}
0 & -z & y \\
z & 0 & -x \\
-y & x & 0
\end{array} \right)
\end{displaymath}
is a skew-symmetric matrix lying in the Lie algebra $\mathfrak{so}(3)$, with the following pro\-perties:
\begin{displaymath}
N^2 = \left( \begin{array}{ccc}
-(y^2 + z^2) & xy & xz \\
xy & -(x^2 + z^2) & yz \\
xz & yz & -(x^2 + y^2)
\end{array} \right)
\end{displaymath}
and $N^3 = -N$, $N^4 = -N^2$, $N^5 = N$, etc.. The matrix $R(\psi, \hat{n})$ arises from geo\-me\-trical
considerations. It is, in fact, the rotation through an angle $\psi$ about an axis along $\hat{n}$.

In fact, using (\ref{RenN}) we can find an angle $\psi$ and a unit vector $\hat{n}$ such that
\[
\left( \begin{array}{ccc}
\cos\phi & -\sin\phi & 0 \\
\sin\phi & \cos\phi & 0 \\
0 & 0 & 1
\end{array} \right) = Id + (\sin \psi) N + (1 - \cos \psi) N^2 .
\]
The equation for the element $R_{33}$ is $1 = 1 - (1 - \cos\psi)(x^2 + y^2)$. Then, $x^2 + y^2 = 0$. Since
this is the sum of two positive numbers, the only solution is $x = y = 0$. This implies that $z^2 = 1$, and
we choose, for reasons to become clear in the next step, $z = -1$. Finally, the equation for the element
$R_{11}$ is $\cos\phi = 1 - (1 - \cos\psi)$; then $\psi = \phi$. The conclusion is
\[
\left( \begin{array}{ccc}
\cos\phi & -\sin\phi & 0 \\
\sin\phi & \cos\phi & 0 \\
0 & 0 & 1
\end{array} \right) = R(\phi, (-1, 0, 0)) .
\]
The corresponding matrix in $SU(2)$ is, then:
\begin{eqnarray}
\left( \begin{array}{ccc}
e^{i\phi/2} & 0 \\
0 & e^{-i\phi/2}
\end{array} \right).
\end{eqnarray}
In the same way, we can also find an angle $\psi$ and a unit vector $\hat{n}$ such that
\[
\left( \begin{array}{ccc}
-\cos\phi & \sin\phi & 0 \\
\sin\phi & \cos\phi & 0 \\
0 & 0 & -1
\end{array} \right) = \mbox{Id} + (\sin \psi) N + (1 - \cos \psi) N^2 .
\]
This matrix to the square is equal to the identity. We can use this fact to find the value of the angle:
\[
\mbox{Id} = \left[\mbox{Id} + (\sin \psi) N + (1 - \cos \psi) N^2\right]^2 = \mbox{Id} + \sin\psi(\sin\psi +
1 - 2\cos\psi) N^2 .
\]
The equality is satisfied if $\sin\psi = 0$, so $\psi = \pi$. The equation for the element $R_{33}$ is $-1 =
1 - 2 (x^2 + y^2)$, and the conclusion is that $x^2 + y^2 = 1$; it follows that $z = 0$. Finally, the
equations for the elements $R_{11}$ and $R_{22}$ are:
\begin{eqnarray}
-\cos\phi &=& 1 - 2 y^2 \\
\sin\phi &=& 1 - 2 x^2 ,
\end{eqnarray}
then we have that $x = \sin(\phi/2)$ and $y = \cos (\phi/2)$. The conclusion is
\[
\left( \begin{array}{ccc}
-\cos\phi & \sin\phi & 0 \\
\sin\phi & \cos\phi & 0 \\
0 & 0 & -1
\end{array} \right) = R(\pi, (\cos(\phi/2), \cos(\phi/2), 0)) .
\]
The corresponding matrix in $SU(2)$ is, then:
\begin{eqnarray}
\left( \begin{array}{ccc}
0 & i e^{i\phi/2}  \\
- i e^{-i\phi/2}  & 0
\end{array} \right).
\end{eqnarray}
From the explicit calculations above, it is now clear that the projective space  can also be written as the
homogeneous space  $SU(2)/H$, where
\begin{displaymath}
H = \left\{ \left( \begin{array}{ccc}
\lambda & 0 \\
0 & \bar{\lambda}
\end{array} \right),
\left( \begin{array}{ccc}
0 & \bar{\lambda} \\
-\lambda & 0
\end{array} \right) | \, \, \, |\lambda|^2 = 1  \right\} ,
\end{displaymath}
is a subgroup of $SU(2)$. In fact, let us note that the orbits of $H$ on $SU(2)$ are generated by terms of
the following form:
\[
(\alpha, \beta) \cdot \lambda = \left( \begin{array}{ccc}
\alpha & \bar{\beta} \\
-\beta & \bar{\alpha}
\end{array} \right)
\left( \begin{array}{ccc}
\lambda & 0 \\
0 & \bar{\lambda}
\end{array} \right) =
\left( \begin{array}{ccc}
\alpha \lambda & \bar{\beta} \bar{\lambda}\\
-\beta \lambda & \bar{\alpha}\bar{\lambda}
\end{array} \right) = (\alpha \lambda, \beta \lambda)
\]
and
\[
(\alpha, \beta) \cdot \lambda = \left( \begin{array}{ccc}
\alpha & \bar{\beta} \\
-\beta & \bar{\alpha}
\end{array} \right)
\left( \begin{array}{ccc}
0 & \bar{\lambda} \\
-\lambda & 0
\end{array} \right) =
\left( \begin{array}{ccc}
-\bar{\beta} \lambda & \alpha \bar{\lambda}\\
-\bar{\alpha} \lambda & -\beta \bar{\lambda}
\end{array} \right) = (-\bar{\beta} \lambda, \bar{\alpha} \lambda).
\]
The set of equivalence classes for this action is clearly equivalent to the quotient $S^2/\mathbb Z_2$, i.e.
to $\bb R P^2$.


\newpage
\begin{thebibliography}{100}
\bibitem{Balachandran1993}
A.~Balachandran, A.~Daughton, Z.~Gu, R.~Sorkin, G.~Marmo, A.~Srivastava, Int.
  J. Mod. Phys. A \textbf{8}, 2993 (1993)

\bibitem{Berry1997}
M.~Berry, J.~Robbins, Proc. R. Soc. London A \textbf{453}, 1771 (1997)

\bibitem{Anastopoulos:02}
C.~Anastopoulos, Int. J. Mod. Phys. A \textbf{19}, 655 (2002)

\bibitem{Kuckert:04}
B.~Kuckert, Physics Letters A \textbf{322}, 47 (2004)

\bibitem{Papadopoulos2004}
N.~Papadopoulos, M.~Paschke, Reyes.~A.F., F.~Scheck, Annales Math\'ematiques
  Blaise-Pascal \textbf{11}(2), 205 (2004). arXiv:quant-ph/0608125v1

\bibitem{Harrison2004}
J.~Harrison, J.~Robbins, J. Math. Phys. \textbf{45}, 1332 (2004)

\bibitem{Peshkin2003}
M.~Peshkin, Phys. Rev. A \textbf{67}, 042102 (2003)

\bibitem{Peshkin2006}
M.~Peshkin, Foundations of Physics \textbf{36}(1), 19 (2006)



\bibitem{Schulman:68}
L.~Schulman, Phys. Rev. \textbf{5}, 1558 (1968)

\bibitem{Laidlaw1971}
M.G. Laidlaw, C.M. DeWitt, Phys. Rev. D \textbf{3}, 1375 (1971)

\bibitem{Leinaas1977}
J.~Leinaas, J.~Myrheim, Il Nuovo Cimento \textbf{37B}, 1 (1977)

\bibitem{Reyes2006}
A.~Reyes, On the Geometry of the Spin-Statistics Connection in Quantum Mechanics. PhD thesis,
 University of Mainz, http://wwwthep.physik.uni-mainz.de/Publications/theses/dis-reyes.pdf (2006)

\bibitem{Benavides2009}
C.~Benavides, A.~Reyes-Lega, in \emph{Geometric and Topological Methods in
  Quantum Field Theory}, ed. by H.~Ocampo, E.~Parigu\'an, S.~Paycha (Cambridge
  Univ. Press, 2009). In press (available at: arXiv:0806.2449v1 [math-ph])

\bibitem{Reyes-Papadopoulos2009}
N.A.~Papadopoulos, A.F.~Reyes-Lega, \emph{On the Geometry of the Berry-Robbins Approach to Spin-Statistics}
(to be published, available at: arXiv:0910.1659v1 [math-ph])

\bibitem{Paschke2001}
M.~Paschke, Von nichtkommutativen Geometrien, ihren Symmetrien und etwas
  Hochenergiephysik.
PhD thesis, University of Mainz, http://wwwthep.physik.uni-mainz.de/Publications/theses/dis-paschke.ps.gz
(2001)

\bibitem{Isham1984}
C.J. Isham, in \emph{Relativity, Groups and Topology II}, ed. by B.S. DeWitt,
  R.~Stora (Kluwer academic publishers, Amsterdam, 1984), pp. 1059--1290


\bibitem{Sudarshan2003}
E.~Sudarshan, I.~Duck, Pramana Journal of Physics \textbf{61}, 1 (2003)

\bibitem{Allen2003}
R.~Allen, A.~Mondragon, Phys. Rev. A \textbf{68}, 046101 (2003)

\bibitem{Streater:homepage}
Streater, R.F.: http://www.mth.kcl.ac.uk/$\sim$streater/lostcauses.html.

\bibitem{Souriau:69}
J.~Souriau, \emph{Structure des Syst\`{e}mes Dynamiques} (Dunod, Paris, 1969)

\bibitem{Woodhouse:80}
N.~Woodhouse, \emph{Geometric Quantization} (Clarendon Press, Oxford, 1980)

\bibitem{Kastrup2003}
H.A. Kastrup, Fortsch.Phys. 51 \textbf{51}, 975 (2003)

\bibitem{Kastrup2006}
H.A.~Kastrup, Annalen der Physik  \textbf{16} 439 (2007)


\bibitem{Kastrup2006a}
H.A.~Kastrup, Physical Review A \textbf{73}, 052104 (2006)

\bibitem{Mackey1968}
G.W. Mackey, \emph{Induced Representations and Quantum Mechanics} (W. A.
  Benjamin, New York, 1968)

\bibitem{Kirillov1976}
A.~Kirillov, \emph{Elements of the Theory of Representations} (Springer Verlag,
  1976)


\bibitem{Sorkin1992}
M.~Bourdeau, R.D.~Sorkin, Phys. Rev. D \textbf{45}, 687 (1992)


\bibitem{Sorkin1992}
M.~Bourdeau, R.D.~Sorkin, Phys. Rev. D \textbf{45}, 687 (1992)

\bibitem{Nab} G. Naber, \textit{Topology, Geometry and Gauge Fields}, New York: Springer-Verlag (1997).

\bibitem{Peshkin2003b}
M.~Peshkin, Phys. Rev. A \textbf{68}, 046102 (2003)

\bibitem{Palais1960}
R.~Palais, Mem. Amer. Math. Soc. \textbf{36} (1960)

\bibitem{Mostow1957}
G.~Mostow, Ann. Math. \textbf{65},432 (1957)

\end{thebibliography}
\end{document}